\documentclass[aps,prd,showpacs,amssymb,amsmath,nofootinbib,preprint]{revtex4}

\usepackage{slashed}

\newcommand{\md}{\mathrm{d}}
\newcommand{\me}{\mathrm{e}}
\newcommand{\mi}{\mathrm{i}}

\newcommand{\mcl}{\mathcal{L}}
\newcommand{\mco}{\mathcal{O}}
\newcommand{\mcs}{\mathcal{S}}
\newcommand{\mbr}{\mathbb{R}}
\newcommand{\mbz}{\mathbb{Z}}
\newcommand{\half}{\tfrac{1}{2}}
\newcommand{\thhalf}{\tfrac{3}{2}}
\newcommand{\quart}{\tfrac{1}{4}}

\DeclareMathOperator{\diag}{diag}

\begin{document}


\title{Kink modes and effective four dimensional fermion and Higgs brane models}
\author{Damien P. George}
\email{d.george@physics.unimelb.edu.au}
\author{Raymond R. Volkas}
\email{r.volkas@physics.unimelb.edu.au}
\affiliation{School of Physics, Research Centre for High Energy
Physics,\\The University of Melbourne, Victoria 3010, Australia}
\date{April 19, 2007}

\begin{abstract}

In the construction of a classical smoothed out brane world model
in five dimensions, one uses a dynamically generated domain wall
(a kink) to localise an effective four dimensional theory.  At the
level of the Euler-Lagrange equations the kink sets up a potential
well, a mechanism which has been employed extensively to obtain localised,
four dimensional, massless chiral fermions.  We present the
generalisation of this kink trapping mechanism for both scalar and
fermionic fields, and retain all degrees of freedom that were
present in the higher dimensional theory.  We show that a kink
background induces a symmetric modified P\"oschl-Teller potential
well, and give explicit analytic forms for all the bound modes and
a restricted set of the continuum modes.  We demonstrate that it
is possible to confine an effective four dimensional scalar field
with a quartic potential of arbitrary shape.  This can be used to
place the standard model electroweak Higgs field on the brane, and
also generate nested kink solutions.  We also consider the limits
of the parameters in the theory which give thin kinks and
localised and de-localised scalar and fermionic fields.

\end{abstract}


\maketitle


\section{Introduction}

Our physical universe is described extremely well by the standard
model and general relativity, both of which are expressed in one
time and three space dimensions.  But these models are incomplete,
and a large amount of work has been done addressing the question:
can we improve our model of the universe by augmenting it with one
or more extra spatial dimensions?  Extra dimensions give extra
degrees of freedom and these have been used to tackle a variety of
problems present in the existing models.

There has been a vast investment of research into serious extra
dimensional models: Universal Extra
Dimensions~\cite{appelquist2000}, the Arkani-Hamed Dimopoulos and
Dvali model~\cite{add,antoniadis1990,aadd} and the Randall-Sundrum
model~\cite{rs1,rs2} to name a few.  The last of these
papers~\cite{rs2} demonstrated that
large, even infinite, extra dimensions are not ruled out by
experiment.  In the large extra dimensional scenarios, one
generically has the concept of a brane, to where the energy which
we observe in our four dimensional world is concentrated, or
confined.  The majority of the Randall-Sundrum (RS) models assume
the existence of a higher theory (e.g. string theory) which
provides such a brane along with a confinement mechanism.  The
effective low energy description of this set-up consists simply of
the combination of a five dimensional bulk action, and a four
dimensional action placed by hand at a specific location in the
extra dimension.  The branes in these models are infinitely thin
and Dirac delta distributions are used to couple the five and four
dimensional sectors.

If our four dimensional universe is embedded in higher dimensions
in such a way, then given enough energy, it should be possible to
probe the structure of our brane.  Theoretical predictions then
require a more sophisticated description of the dynamics of the
brane than just a delta distribution.  Ideally, one would like
to give a dynamical explanation for the formation of a brane, and
the ensuing localisation, using classical field theoretic ideas.
The canonical example of this comes from the work done initially
by Rubakov and Shaposhnikov in~\cite{rubakov1983} (see
also~\cite{akama1982,visser1985}).  One has a five
dimensional scalar field $\Phi(x^\mu,w)$ in a potential with
$\mbz_2$ symmetry that admits a topologically stable solution
$\phi_c(w) \sim \tanh (m w)$.  The mass parameter $m$ controls the
tension or inverse width of this domain wall defect.  A massless
five dimensional
fermion with a Yukawa coupling to this kink has a Dirac equation
which admits a separated solution
$\Psi(x^\mu,w) \sim \psi_L(x^\mu) \cosh^{-a}(m w)$. The parameter
$a$ is proportional to the five dimensional Yukawa coupling
constant and $\psi_L$
is a massless left-handed four dimensional fermion.  In the
limit of a thin brane, the extra dimensional factor of $\Psi$
becomes a delta distribution, giving essentially the RS
localisation mechanism, but with known dynamical origins.

Certainly, if one is just interested in the exact thin brane
limit, then all degrees of freedom freeze out except for the
massless zero mode $\psi_L$.  But if the brane is dynamically
generated, this limit will only be approximate, and at high enough
energies the excited states (essentially the Kaluza-Klein modes
of the infinite extra dimension) will become phenomenologically
important.  Gaining an understanding of the behaviour of these
excitations is therefore a prerequisite for any realistic model
building attempts.

Even with a full grasp of the dynamics of the confinement of
massless chiral modes, one is still not in a position to
dynamically embed the standard model in an extra dimension.  In
any realistic interacting model, there will be gauge symmetries
that need to be broken, most notably the electroweak symmetry.
The most straightforward idea is to localise the standard Higgs
field to the brane, and then proceed in the usual manner of
spontaneous symmetry breaking.  Such a mechanism requires an
analysis of the five dimensional couplings needed to produce an
effective four dimensional quartic potential, and a compatible
way of confining gauge fields.

In this paper we address some of these model building issues by
demonstrating that it is possible to confine an effective Higgs
field to a kink.  We also present the full spectrum of modes of
the kink, a coupled scalar field and a coupled fermion field in
the canonical kink scenario.  Along with this mode analysis, we
also investigate the relevant limits of the parameters in the
model which yield a thin kink, and show that we can obtain just
what is needed for an embedding of the standard model.  We do
not address the issue of gauge field confinement; one promising
idea is the Dvali-Shifman mechanism~\cite{dvali1996}.  Gravity
will also be ignored for the sake of simplicity in this initial
pass through the problem.

We begin in Section~\ref{sec:kink} by presenting the toy model
which supports a scalar kink and determine the full spectrum of
its associated modes.  We discuss the different limits of this
model which give the scenarios with no kink, a thick kink and a
thin kink.  In Section~\ref{sec:scalar} we add a scalar field to
the kink model, and show that the kink sets up a symmetric
modified P\"oschl-Teller potential for the extra dimensional
component of the scalar field.  We determine the modes of this
potential and use them to obtain an effective four dimensional
action, discussing in detail the thin kink limit.  In
Section~\ref{sec:fermion} we analyse a fermion coupled to the
kink, present the full mode decomposition, and show that in the
thin kink limit, the massless left-handed mode is the only
surviving dynamical field.  We also present an action that
contains this massless four dimensional mode coupled to a five
dimensional field.  We conclude and discuss further work in
Section~\ref{sec:conc}.  Appendix~\ref{app:smpt} contains
analytic solutions of the potential well set up by the kink which
are used extensively throughout the analysis.
Appendix~\ref{app:freeze} details the dynamics of the translation
symmetry of the kink, and we show how the zero mode of translation
freezes out in the thin kink limit.


\section{The kink and its limits}
\label{sec:kink}

All subsequent work will be in $4+1$ dimensional Minkowski space
with metric $g_{M N} = \diag (1,-1,-1,-1,-1)$.  Capital Latin
letters index the full space, Greek letters index the $3+1$
dimensional subspace, the extra dimensional coordinate is $w$ and
the embedding is $x^M = (x^\mu,w)$.
We begin with an action
describing a real five dimensional scalar field $\Phi$, given by
\begin{align}
\label{eq:phi-act-5d}
\mcs_\Phi &= \int \md^5 x \left[ \half \partial^M \Phi \partial_M \Phi - V(\Phi) \right] \\
& \text{with} \quad V(\Phi) = \frac{a}{4m} \left( \Phi^2 - \frac{m^3}{a} \right)^2,
\end{align}
where $a$ is a dimensionless constant\footnote{The constant $a$
is the five dimensional analogue of $\lambda$ in the four
dimensional potential
$V(\phi)=\tfrac{\lambda}{4}(\phi^2-\tfrac{m^2}{\lambda})^2$.}
and $m$ is the mass of $\Phi$.  From this action we find the
Euler-Lagrange equation for $\Phi$ to be
\begin{equation}
\label{eq:phi-kink}
\partial^M \partial_M \Phi - m^2 \Phi + \frac{a}{m} \Phi^3 = 0.
\end{equation}

With the aim of producing an effective four dimensional theory, 
$\Phi$ will initially be taken to depend only on the extra
dimensional coordinate $w$; this behaviour is denoted as
$\phi_c(w)$.  A topologically stable solution to
Eq.~\eqref{eq:phi-kink} is then
\begin{equation}
\label{eq:phi-clas}
\phi_c(w) = \sqrt\frac{m^3}{a} \tanh \left( \frac{mw}{\sqrt2} \right),
\end{equation}
which is the classical kink solution interpolating between the
$\mbz_2$ degenerate minima of the potential $V$.  This kink has
constant energy per unit volume at every spatial point in the
$3+1$ dimensional subspace, given by
\begin{equation}
\label{eq:phi-energy}
\varepsilon_{\phi_c}
    = \int_{-\infty}^{\infty} \md w \left[ \half (\partial_w \phi_c)^2 + V(\phi_c) \right]
    = \frac{2\sqrt2 m^4}{3 a}.
\end{equation}

We are interested in the behaviour of models where the classical
background is a thin kink, meaning $m$ is very large.  To make
this more precise, we write our parameters as
\begin{align*}
a &= \tilde{a} \Lambda^\alpha, &
m &= \tilde{m} \Lambda^\mu,
\end{align*}
with $\tilde{a}$ and $\tilde{m}$ finite, and consider the limit
$\Lambda \rightarrow \infty$.  In such a limit the kink energy
density is $\varepsilon_{\phi_c} \sim \Lambda^{4\mu-\alpha}$
and must remain finite, giving the constraint $\alpha = 4\mu$.
This in turn means the amplitude of the kink is
$|\phi_c| \sim \Lambda^{-\half\mu}$.  The single parameter $\mu$
now describes all possible limiting scenarios of the theory, with
$\mu=0$ corresponding to no limit being taken.  If $\mu<0$ the
potential $V$ disappears, $\phi_c \rightarrow 0$\footnote{The
relevant limit is
$\lim_{m\rightarrow0} \tanh(mw/\sqrt2) / \sqrt{m}= 0$.}
and the action describes a massless, freely propagating, five
dimensional scalar field.  The case $\mu>0$ is the
more interesting thin kink limit.  Here, the width of $\phi_c$
tends to zero and to keep the energy density finite, the height
also vanishes.  We will refer extensively to these limits in the
following sections.

By assuming that $\Phi$ depends only on $w$ we have of course lost
a lot of the dynamics of the full theory.  First, since $\phi_c$
breaks translational invariance along $w$, we expect a zero mode
which can act to translate the kink.  Second, if we have a thick
kink, we expect there to be massive modes associated with
arbitrary deformations of the kink.  We now proceed to incorporate
these dynamics.


\subsection{Kink modes and the effective model}

The classical kink background $\phi_c$ breaks the five dimensional
Poincar\'e symmetry, leaving a four dimensional Poincar\'e
subgroup.  This makes it natural to decompose a field into a sum
of products of an extra dimensional component and a $3+1$
dimensional component.  For $\Phi$, we want this expansion to be
made about the kink solution, and so we take
\begin{equation}
\label{eq:phi-expand}
\Phi(x^\mu,w) = \phi_c(w) + \sum_i \phi_i(x^\mu) \eta_i(w),
\end{equation}
where $\eta_i$ are a fixed orthonormal basis of the extra
dimension, $\phi_i$ are four dimensional dynamical fields and the
sum over $i$ can in general be a combination of discrete and
continuous modes.  We would like to determine a basis $\eta_i$
such that the equations of motion for the $\phi_i$ describe
massive scalar fields.  This can be done in the standard way by
taking the action for $\Phi$ given by~\eqref{eq:phi-act-5d},
substituting the expansion~\eqref{eq:phi-expand}, using the fact
that $\phi_c$ satisfies~\eqref{eq:phi-kink}, discarding terms
$\mco(\phi_i \eta_i)^3$ and higher and using integration by parts.
The effective second order action is then
\begin{equation*}
\mcs_\Phi^{\mco(2)} = \int \md^5 x \left[
    \frac{a}{4m} \phi_c^4 - \frac{m^5}{4a}
    - \half \phi_i \eta_i \left( \partial^\mu \partial_\mu - \partial_w^2
                                 + \frac{3a}{m} \phi_c^2 - m^2 \right) \phi_j \eta_j
    \right],
\end{equation*}
with implicit sum $i,j$ over the modes.  For the $\phi_i$ to
satisfy the massive Klein-Gordon equation with mass $\lambda_i$ we
require
\begin{equation*}
\left( -\frac{\md^2}{\md w^2} + \frac{3a}{m} \phi_c^2 - m^2 \right) \eta_i = \lambda_i^2 \eta_i.
\end{equation*}
We can use the known form of $\phi_c$ to get
\begin{equation}
\label{eq:eta-defn}
\left( -\frac{\md^2}{\md z^2} + 6 \tanh^2 z - 2 \right) \eta_i = \left( \frac{2 \lambda_i^2}{m^2} \right) \eta_i,
\end{equation}
where $z=mw/\sqrt2$.  This differential equation is a
Schr\"odinger equation with a symmetric modified P\"oschl-Teller
potential.  Analytic solutions are known in terms of
hypergeometric functions, and in general there are both bound and
continuum solutions.  In Appendix~\ref{app:smpt} we present these
solutions expressed in terms of regular functions, along with
their normalisation coefficients, for a more general form of the
potential.  For the kink modes at hand,
Eq.~\eqref{eq:eta-defn} is Eq.~\eqref{eq:smpt} with
$l=2$, and so there are two bound modes and a continuum
\begin{align*}
\lambda_0^2 &= 0 &
    \eta_0(w) &= E_0 \cosh^{-2} z, \\
\lambda_1^2 &= \thhalf m^2 &
    \eta_1(w) &= E_1 \sinh z \cosh^{-2} z, \\
\lambda_q^2 &= \half (q^2 + 4) m^2 &
    \eta_q(w) &= E_q \me^{\mi q z} \left( 3 \tanh^2 z - (q^2+1) - 3 \mi q \tanh z \right).
\end{align*}
The bound $\eta_{0,1}$ are square integrable normalised
by~\eqref{eq:norm-sqint} and the continuum $\eta_q$ are delta
function normalised by~\eqref{eq:norm-delta}, the normalisation
constants being
\begin{align*}
E_0 &= \sqrt\frac{3m}{4\sqrt{2}}, &
E_1 &= \sqrt\frac{3m}{2\sqrt{2}}, &
E_q &= \sqrt{\frac{m}{2 \pi \sqrt2 (q^2+1)(q^2+4)}}.
\end{align*}

Armed with the basis $\eta_i$, we return to the analysis of the
full dynamics of the kink.  Expanding the original
action~\eqref{eq:phi-act-5d} with $\Phi$ decomposed in the
$\eta_i$ basis and integrating over the extra dimension gives
\begin{equation}
\label{eq:phi-act-4d}
\mcs_\Phi = \int \md^4 x \left[ -\varepsilon_{\phi_c} + \mcl_\phi \right],
\end{equation}
where the $\phi$ kinetic, mass and self coupling terms are
\begin{equation}
\label{eq:phi-lag-4d}
\begin{aligned}
\mcl_\phi &=
    \half \partial^\mu \phi_0 \partial_\mu \phi_0
    + \half \partial^\mu \phi_1 \partial_\mu \phi_1
    - \tfrac{3}{4} m^2 \phi_1^2 \\
& \quad
    + \int_{-\infty}^{\infty} \md q \left[
        \half \partial^\mu \phi_q^* \partial_\mu \phi_q
        - \quart (q^2 + 4) m^2 \phi_q^* \phi_q \right] \\
& \quad
    - \kappa^{(3)}_{i j k} \phi_i \phi_j \phi_k
    - \kappa^{(4)}_{i j k l} \phi_i \phi_j \phi_k \phi_l.
\end{aligned}
\end{equation}
The fields $\phi_{0,1}$ are the real valued scalars associated
with the two bound state modes $\eta_{0,1}$.  The integration over
$q$ is over the complex valued continuum modes $\phi_q$ associated
with $\eta_q$.  Note that $\phi_{-q}=\phi_q^*$ and
$\eta_{-q}=\eta_q^*$ and so this integral is real.  The effective
cubic and quartic self interaction couplings are
\begin{align*}
\kappa^{(3)}_{i j k} &= \frac{a}{m} \int_{-\infty}^\infty \md w \left[ \phi_c \eta_i \eta_j \eta_k \right], \\
\kappa^{(4)}_{i j k l} &= \frac{a}{4m} \int_{-\infty}^\infty \md w \left[ \eta_i \eta_j \eta_k \eta_l \right].
\end{align*}
For brevity, the indices $i,j,k,l$ label bound modes, continuum
modes or a mixture of both and the sum over these labels is
implied in Eq.~\eqref{eq:phi-lag-4d}.  The couplings $\kappa$
can be computed as their integrands are known; some are zero due
to parity, some are non-zero.

Equations~\eqref{eq:phi-act-4d} and~\eqref{eq:phi-lag-4d} are
exact manipulations of the original five dimensional $\Phi$
model~\eqref{eq:phi-act-5d}, and provide a description in a
basis useful for investigating the effective four dimensional
behaviour.  The bound modes are reminiscent of the Kaluza-Klein
modes one obtains in compact extra dimensions, but in the case at
hand the gaps in the mass spectrum are not uniform.  Furthermore,
the continuum modes do not have analogues in the Kaluza-Klein
model and are not strictly four dimensional, but instead form a
pseudo-five dimensional field with reduced degrees of freedom.

Regarding the renormalisability of our model, we note that because
the original five dimensional action~\eqref{eq:phi-act-5d}
contains non-renormalisable terms, we do not have any reason
to stop writing down potentials at quartic order.  But the
manipulations which bring us to the four dimensional level given
by Eq.~\eqref{eq:phi-lag-4d}, do in fact leave us with a
renormalisable theory if we truncate the action to just the bound
states.  This renormalisability of the bound states may be a
useful criterion for restricting the types of terms that one begins
with in the five dimensional (or higher dimensional) action.


\subsection{Limiting behaviour of the effective model}

Now that we have such a reformulation of the kink model, we are in
a position to analyse the full dynamics of the system for the
three different limits of the mass $m$.  Following our previous
parameterisation, for $\mu<0$ there is no kink and the basis
$\eta_i$ is replaced by a standard complex Fourier expansion.  All
the dynamical components are packaged together by the Fourier
transform and it is no longer sensible to perform the $w$
integral.  Instead one should consider $\Phi$ as a free, massless,
five dimensional field.

For the thick kink case when $\mu=0$, the spectrum consists of the
energy density of the integrated kink, a zero mode, a massive
bound mode and a continuum of massive complex scalars.  These
effective four dimensional fields are self-coupled and coupled
amongst each other via cubic and quartic interactions.  In
particular, the zero mode $\phi_0$ and massive bound mode $\phi_1$
each have a potential:
\begin{align*}
V_0(\phi_0) &= \frac{9 \sqrt2 \, a}{140} \phi_0^4, \\
V_1(\phi_1) &= \thhalf m^2 \phi_1^2
               + \frac{3\pi}{32} \sqrt{\frac{3a}{2\sqrt2}} \, m \, \phi_1^3
               + \frac{9\sqrt2 \, a}{280} \phi_1^4,
\end{align*}
which are due to the non-zero values of $\kappa^{(4)}_{0000}$,
$\kappa^{(3)}_{111}$ and $\kappa^{(4)}_{1111}$.  Similarly, we can
compute the coupling potential amongst these bound modes, to get
\begin{equation*}
V_{0,1}(\phi_0,\phi_1) =
    \frac{9\pi}{64} \sqrt{\frac{3a}{2\sqrt2}} \, m \, \phi_0^2 \phi_1
    + \frac{9\sqrt2 \, a}{70} \, \phi_0^2 \phi_1^2.
\end{equation*}

Note that while $\phi_0$ has no mass term it does have a non-zero
potential, making it energetically unfavourable to excite the
field, even though it costs zero energy to translate the kink.  We
can account for this unexpected result by recalling that
$\eta_0$ corresponds to \emph{infinitesimal} translations of
$\phi_c$.  Adding any small but finite multiple of $\eta_0$
to $\phi_c$ will, to first order, perform a translation, but to
higher order it will \emph{deform} the kink.  The energy cost of
these higher order deformations are described by the potential
$V_0$.  For completeness we point out that while $V_1$ has a cubic
term, the potential has only one extremum, which is a minimum at
$\phi_1=0$.

We now move on to the thin kink limit where $\mu>0$ and both $m$
and $a$ tend to infinity.  The kink energy density remains finite,
but the masses of the bound mode $\phi_1$ and the continuum modes
$\phi_q$ go like $m$.  Recall that these massive modes correspond
to deformations in the kink and as the kink gets thinner it also
gets stiffer, requiring more energy for a given deformation.  The
dynamics of the massive modes are thus frozen out, as they are no
longer able to deform the kink without possessing infinite energy.
This argument also applies to the translation zero mode $\phi_0$,
which gets frozen out in the thin kink limit, despite remaining
massless, because its
corresponding potential $V_0$ becomes infinitely steep.  This
can be understood from the arguments give above: for an infinitely
stiff kink, the higher order deformations due to the zero mode are
forbidden and the only physical resolution is to remove the
dynamics of this mode.  So in the thin kink limit, we are left
with only the kink energy density $\varepsilon_{\phi_c}$ in the
effective four dimensional action.  Appendix~\ref{app:freeze}
discusses this result in more depth and includes an identification
of the mode responsible for finite translations.  Also
see~\cite{shaposhnikov2005} for a discussion of the translational
zero mode in the presence of gravity.


\section{Adding a scalar field}
\label{sec:scalar}

We have so far performed an analysis of the kink and its modes in
isolation.  As stated previously, we aim to use the properties of
the domain wall to dynamically trap five dimensional fields to a
brane and create an effective four dimensional model.  We can
achieve this if the kink is coupled to a different five
dimensional field and projects out a set of modes with the lowest
mode separated from the rest by a significant mass gap.  Then if
we are at energies where only the lowest bound state can be
excited, the degree of freedom of propagation along the extra
dimension has been lost and the bound mode is confined.


\subsection{Scalar modes}

The simplest place to start is to take another five dimensional
scalar field $\Xi(x^M)$ and couple it to the kink field.  The
action describing this model is
$\mcs_{\Phi+\Xi} = \mcs_\Phi + \mcs_\Xi$ where
\begin{align}
\label{eq:xi-act-5d}
\mcs_\Xi &= \int \md^5 x \left[
    \half \partial^M \Xi \partial_M \Xi
    - \frac{a b(b+1)}{4 m} \Phi^2 \Xi^2
    - W(\Xi)
    \right] \\
& \text{with} \quad W(\Xi) = \frac{n^2}{2} \Xi^2 + \frac{c}{4n} \Xi^4.
\end{align}
The parameters $b$ and $c$ are dimensionless, while the
dimensionful parameter $n$ is the mass of $\Xi$.  We follow the
analysis for the kink modes and perform the general separation
\begin{equation*}
\Xi(x^M) = \sum_i \xi_i(x^\mu) k_i(w),
\end{equation*}
where the sum over $i$ can again be a combination of discrete and
continuous parts.  To obtain a suitable basis $k_i$, we look at
the linearised equation of motion for $\Xi$ with $\Phi = \phi_c$
\begin{equation*}
\left( \partial^\mu \partial_\mu - \partial_w^2 + \frac{a b(b+1)}{2 m} \phi_c^2 + n^2 \right) \xi_i k_i = 0.
\end{equation*}
We want $\xi_i$ to satisfy the four dimensional Klein-Gordon
equation with mass $\delta_i$.  This leads to
\begin{equation*}
\left( -\frac{\md^2}{\md z^2} + b(b+1) \tanh^2 z \right) k_i = \left( \frac{2(\delta_i^2 - n^2)}{m^2} \right) k_i
\end{equation*}
where $z=mw/\sqrt2$ as before.  This Schr\"odinger equation has the same
form as the one obtained for the kink modes.  We see that the kink
sets up a symmetric modified P\"oschl-Teller potential well which
traps not only its own modes, but also those of a coupled scalar
field.  Looking to Appendix~\ref{app:smpt} we see that the basis
$k_i$ contains $\lceil b \rceil$ bound modes\footnote{We use the
standard notation $\lceil . \rceil$ for the ceiling function.} and
a continuum.  The masses of the bound states are
\begin{align*}
\delta_0^2 &= n^2 + \half b m^2, \\
\delta_1^2 &= n^2 + \half (3b-1) m^2, \\
\delta_2^2 &= n^2 + \half (5b-4) m^2, \\
\vdots \\
\delta_i^2 &= n^2 + \half ((2i+1)b-i^2) m^2,
\end{align*}
and for the continuum we have
\begin{align*}
\delta_q^2 &= n^2 + \half (q^2 + b(b+1)) m^2,
\end{align*}
where $q\in\mbr$ labels the continuum modes.  We will not give
explicit forms of the functions $k_i$; they are easily determined
from Appendix~\ref{app:smpt}.  Unlike the modes of the kink, this
spectrum of masses does not in general include a zero mode and the
bottom of the spectrum is dependent on the parameters $n$, $b$ and
$m$.  We also have the freedom to change the sign of $n^2$ in the
original action and dial up any positive, zero, or negative value
of $\delta_0^2$.

As before, we use the basis $k_i$ to expand $\Xi$ in the original
action~\eqref{eq:xi-act-5d} and integrate over the extra
dimension.  Including the kink sector, the effective four
dimensional action is then
\begin{equation*}
\mcs_{\Phi+\Xi} = \int \md^4 x \left[ -\varepsilon_{\phi_c} + \mcl_\phi + \mcl_\xi \right],
\end{equation*}
where the kink-only parts are given previously and the scalar
Lagrangian is
\begin{equation}
\label{eq:xi-lag-4d}
\begin{aligned}
\mcl_\xi &=
    \sum_{i=0}^{\lceil b-1 \rceil} \left[ \half \partial^\mu \xi_i \partial_\mu \xi_i
    - \half \delta_i^2 \xi_i^2 \right]
    + \int_{-\infty}^{\infty} \md q \left[ \half \partial^\mu \xi_q \partial_\mu \xi_q
    - \half \delta_q^2 \xi_q^2 \right] \\
& \quad
    - g^{(3)}_{ijk} \phi_i \xi_j \xi_k
    - g^{(4)}_{ijkl} \phi_i \phi_j \xi_k \xi_l
    - \tau_{ijkl} \xi_i \xi_j \xi_k \xi_l.
\end{aligned}
\end{equation}
The Yukawa and self coupling factors are
\begin{align*}
g^{(3)}_{ijk}  &= \frac{ab(b+1)}{2m} \int_{-\infty}^\infty \md w \left[ \phi_c \eta_i k_j k_k \right], \\
g^{(4)}_{ijkl} &= \frac{ab(b+1)}{4m} \int_{-\infty}^\infty \md w \left[ \eta_i \eta_j k_k k_l \right], \\
\tau_{ijkl}    &= \frac{c}{4n}       \int_{-\infty}^\infty \md w \left[ k_i k_j k_k k_l \right].
\end{align*}

With this expanded four dimensional action, we are ready to
analyse the various limits of the model with the scalar field.


\subsection{The thin kink with a scalar field}

We are considering the combined five dimensional action
$\mcs_{\Phi+\Xi} = \mcs_\Phi + \mcs_\Xi$ and the various limits
that arise through $m$, the mass of the kink.  As discussed
previously, we have three scenarios which are characterised by the
sign of $\mu$.  In the case $\mu<0$ there is no kink and we are
left with two interacting five dimensional fields $\Phi$ and
$\Xi$.  The thick kink scenario, $\mu=0$, has many interacting
four dimensional scalar fields, the details given by the mass
spectra of $\phi_i$ and $\xi_i$ and the couplings $\kappa$, $g$
and $\tau$.  We will not dwell on these two cases, but instead
concentrate our attention on the thin kink limit and explore the
parameter space of $b$, $c$ and $n$.

It was shown previously that the thin kink limit leaves only the
energy density of the domain wall in the effective four
dimensional action.  The dynamics of all the scalar modes $\phi_i$
are removed and so the $g^{(3)}$ and $g^{(4)}$ Yukawa terms
in~\eqref{eq:xi-lag-4d} are eliminated\footnote{It is shown in
Appendix~\ref{app:freeze} that the translation symmetry manifests
as a linear combination of $\phi_{0,1,q}$ which can assume a
non-zero constant value with zero energy cost, even in the thin
kink limit.  It is possible to show that this contribution to the
Yukawa terms is counteracted by shifting the $k_i$ basis to align
with the shifted kink profile, and redefining the fields
$\xi_i$.}.  With $m\rightarrow\infty$
and $b$ finite, the masses of all the $\xi_i$ modes will also tend
to infinity and the scalar $\Xi$ becomes completely frozen out.
To leave some remnant of $\Xi$ in the model we have two choices:
either take $b$ to zero to counter $m^2$, or choose $n^2$ such
that it cancels $\half b m^2$.

For the first choice, let $n$ be finite and
$b=\tilde b \Lambda^\beta$.  Then
$\delta_0^2 \sim \Lambda^{\beta+2\mu}$ and the mode $\xi_0$ has
finite mass if $\beta+2\mu\le0$.  Since $b\rightarrow0$ in the
limit $\Lambda\rightarrow\infty$, there are in fact no bound
modes, the basis $k_i$ is not valid and we must consider $\Xi$ as
a five dimensional field.  The effective action for such a limit
of the parameters is\footnote{This result uses
$\tanh^2(mw/\sqrt2) \rightarrow 1$ as $m\rightarrow\infty$, which
ignores the fact that the distribution vanishes on a set of
measure zero at the origin.}
\begin{equation}
\label{eq:xi-eff-5d}
\mcs_{\Phi+\Xi}^\text{5D} = \int \md^4 x \left[ -\varepsilon_{\phi_c} \right]
    + \int \md^5 x \left[ \half \partial^M \Xi \partial_M \Xi - \quart b m^2 \Xi^2 - W(\Xi) \right].
\end{equation}
We see that the five dimensional field $\Xi$ has nothing dynamical
to couple to, and just picks up an addition to its $\Xi^2$ term.
If we have the strict inequality $\beta+2\mu<0$, this addition to
the mass will be zero.

The second choice which keeps some part of $\Xi$ alive is to
change the sign of $n^2$ and fine tune it to exactly cancel the
infinite term in $\delta_0^2$.  An exact cancellation would
render the mode $\xi_0$ massless; we can be more general and allow
a finite mass to remain by choosing
\begin{equation}
\label{eq:nsq-cancel}
n^2 = -\half b m^2 + n_0^2,
\end{equation}
where $n_0$ is finite.  Let us briefly comment on the situation
where $b\rightarrow0$, but not quickly enough to counter $m$
(thus $-2\mu<\beta<0$) and so we must choose $n^2$ as
in~\eqref{eq:nsq-cancel}.  In this case there are again no bound
modes and $\Xi$ is a five dimensional field with equivalent
physics as described by~\eqref{eq:xi-eff-5d}, except the
quadratic term of the generated potential is replaced by
$\half n_0^2 \Xi^2$.

We can now restrict our analysis to the case where $\beta\ge0$
and $n^2$ is of the form given by~\eqref{eq:nsq-cancel}.  As we
have a non-zero $b$, there is at least one bound mode, and in fact
only the lowest bound mode will have finite mass.  The dynamics of
the higher bound modes and the continuum will not be part of the
model as their corresponding masses are infinite.  With only
$\xi_0$ alive, the quartic coupling terms reduce to just the
one with the factor $\tau_{0000}$.  Putting all these pieces
together we arrive at the four dimensional action
\begin{align*}
\mcs_{\Phi+\Xi}^\text{4D} &= \int \md^4 x \left[
    -\varepsilon_{\phi_c}
    + \half \partial^\mu \xi_0 \partial_\mu \xi_0
    - W_0(\xi_0)
    \right] \\
& \text{with} \quad W_0(\xi_0) =
    \half n_0^2 \xi_0^2
    + \frac{c}{4\sqrt{2\pi}}
        \frac{\sqrt2 \, \Gamma^2(b+\half)\Gamma(2b)}{\sqrt b \, \Gamma^2(b)\Gamma(2b+\half)}
        \xi_0^4.
\end{align*}
For large values of $b$ the potential simplifies to
\begin{equation*}
W_0(\xi_0) \sim \half n_0^2 \xi_0^2 + \frac{c}{4\sqrt{2\pi}} \xi_0^4.
\end{equation*}

This analysis shows that in the thin kink limit, a five
dimensional coupled scalar field is projected down to a single,
localised, four dimensional scalar field $\xi_0$.  As the
parameters $n_0$ and $c$ are arbitrary, one can generate a
phenomenologically suitable potential for $\xi_0$, in particular
the sign of $n_0^2$ can be changed to yield a potential which
encourages a non-zero vacuum expectation value.  We mention two
uses of this.  Most obviously this mechanism can be used to
localise the standard model electroweak Higgs field to a brane, and it should
be no trouble to arrange Yukawa couplings to fermion fields for
mass generation.  Secondly, the effective potential $W_0(\xi_0)$
has the same form as the kink potential $V(\Phi)$ and can thus
support a domain wall solution, leading to the idea of nested
brane worlds.  Beginning with a six dimensional model with the two
scalar fields $\Phi$ and $\Xi$ and a suitable potential, one can
use $\Phi$ to generate a domain wall and an effective five
dimensional action and then use the lowest projected mode of $\Xi$
in the same way to generate an effective four dimensional action.
The method used does not depend on the dimensionality and one is
free to generate an arbitrary number of nestings.  Of course, this
mechanism just deals with the particle content.  The non-trivial
exercise is to check that gravity can be broken down in similar
stages and made to reproduce four dimensional general relativity;
this will not be attempted here.


\section{Adding a fermion field}
\label{sec:fermion}

In the previous section we performed a full analysis of the modes
of a five dimensional scalar field coupled to a kink.  We showed
that in the thin kink limit, an effective four dimensional scalar
field remains and could be potentially useful for model building.
In direct analogy with this analysis we now consider a five
dimensional massless fermion with Yukawa coupling to $\Phi$, find
a suitable basis for decomposition, and investigate the limiting
behaviour.  This is a generalisation of the well known and heavily
used result that the kink supports a zero mode fermion, as first
discussed in the brane context in~\cite{rubakov1983}.  For a
partially analytic analysis of massive fermion modes confined to a
thick brane in the presence of gravity, see~\cite{ringeval2002}.


\subsection{Fermion modes}

Fermions in five dimensions are four component spinors and their
Dirac structure is described by $\Gamma^M$ with
$\{\Gamma^M,\Gamma^N\}=g^{M N}$.  Specifically
$\Gamma^\mu = \gamma^\mu$ and $\Gamma^5=\mi \gamma^5$ where
$\gamma^{\mu,5}$ are the usual gamma matrices in the Dirac
representation.  Our action for a massless fermion coupled to the
kink is $\mcs_{\Phi+\Psi} = \mcs_\Phi + \mcs_\Psi$ where
\begin{equation}
\label{eq:psi-act-5d}
\mcs_\Psi = \int \md^5 x \left[
    \overline\Psi \mi \Gamma^M \partial_M \Psi
    - \sqrt{\frac{ad^2}{2m}} \Phi \overline\Psi \Psi
    \right].
\end{equation}
The kink parameters $a$ and $m$ are the same as before and $d$ is
a dimensionless coupling parameter.  As with the scalar field
$\Xi$, we expect the extra dimensional behaviour of $\Psi$ to be
quite different to the four dimensional part.  Also, because of
the Dirac structure of the fifth gamma matrix
$\Gamma^5 = \mi \gamma^5$, we expect left- and right-handed
projections of the four dimensional part to behave differently.
Thus we choose the general expansion
\begin{equation}
\label{eq:psi-expand}
\Psi(x^\mu,w) = \sum_i \psi_{L i}(x^\mu) f_{L i}(w)
              + \sum_i \psi_{R i}(x^\mu) f_{R i}(w),
\end{equation}
where the $f_{L i}$ and $f_{R i}$ are a fixed basis, the $\psi_i$
are dynamical, $\gamma^5 \psi_{L i} = -\psi_{L i}$,
$\gamma^5 \psi_{R i} = \psi_{R i}$ and the sum over $i$ can be
both discrete and continuous.  To obtain the defining equations
for the basis functions $f_i$, we impose that the $\psi_i$ satisfy
the massive Dirac equation by
$\mi \slashed\partial \psi_{L i} = \sigma_i \psi_{R i}$
and $\mi \slashed\partial \psi_{R i} = \sigma_i \psi_{L i}$.
Then, substituting the expansion~\eqref{eq:psi-expand} into the
Euler-Lagrange equation for $\Psi$ we arrive at
\begin{equation}
\begin{aligned}
& \psi_{L i} \left( -\partial_w f_{L i} + f_{R i} \sigma_i - \sqrt{\frac{ad^2}{2m}} \phi_c f_{L i} \right) \\
& + \psi_{R i} \left( \partial_w f_{R i} + f_{L i} \sigma_i - \sqrt{\frac{ad^2}{2m}} \phi_c f_{R i} \right)
    = 0.
\end{aligned}
\label{eq:f-defn1}
\end{equation}

Since left and right Dirac components are independent and the
$\psi_i$ are arbitrary fields, both of the two parenthesised
factors in Eq.~\eqref{eq:f-defn1} must be zero.  Hence the
$f_{L i}$ and $f_{R i}$ must satisfy a set of two first order
coupled ordinary differential equations.  We turn these equations
into two uncoupled second order equations:
\begin{align*}
\left( -\frac{\md^2}{\md z^2} + d(d+1) \tanh^2 z - d \right) f_{L i} &= \frac{2 \sigma_i^2}{m^2} f_{L i}, \\
\left( -\frac{\md^2}{\md z^2} + d(d-1) \tanh^2 z + d \right) f_{R i} &= \frac{2 \sigma_i^2}{m^2} f_{R i}.
\end{align*}
As with the scalar field, we see that the kink sets up a symmetric
modified P\"oschl-Teller potential well which traps the extra
dimensional component of the fermion field.  This has been noted
in a similar context of fat branes in~\cite{hung2004}, and in the
context of a two dimensional Dirac equation
in~\cite{decastro2006}.  We use Appendix~\ref{app:smpt} of the
current work to obtain the solutions; the bound modes come in
pairs given by
\begin{align*}
(\sigma_0^d)^2 &= 0 && \Biggl\{
    \begin{aligned}
        f_{L 0}^d(w) &= F_{L 0}^d \cosh^{-d} z \\
        f_{R 0}^d(w) &= 0,
    \end{aligned} \\
(\sigma_1^d)^2 &= \half (2d-1) m^2 && \Biggl\{
    \begin{aligned}
        f_{L 1}^d(w) &= F_{L 1}^d \sinh z \cosh^{-d} z \\
        f_{R 1}^d(w) &= F_{R 1}^d \cosh^{-d+1} z,
    \end{aligned} \\
\vdots \\
(\sigma_n^d)^2 &= \half (2nd-n^2) m^2 && \Biggl\{
    \begin{aligned}
        f_{L n}^d(w) &= \frac{1}{\sigma_n^d} \left( \frac{dm}{\sqrt2} \tanh z - \frac{\md}{\md w} \right) f_{L, n-1}^{d-1} \\
        f_{R n}^d(w) &= f_{L, n-1}^{d-1}(w).
    \end{aligned} \\
\end{align*}
These bound state modes are valid for all positive values of
$d$ and there are $\lceil d \rceil$ sets of modes.  For the
continuum, the solutions can be found in terms of standard
functions when $d$ is a positive integer.  They are
\begin{align*}
(\sigma_q^1)^2 &= \half (q^2 + 1) m^2 && \Biggl\{
    \begin{aligned}
        f_{L q}^1 &= F_{L q}^1 \me^{\mi q z} \left( \tanh z - \mi q \right) \\
        f_{R q}^1 &= F_{R q}^1 \me^{\mi q z},
    \end{aligned} \\
(\sigma_q^2)^2 &= \half (q^2 + 4) m^2 && \Biggl\{
    \begin{aligned}
        f_{L q}^2 &= F_{L q}^2 \me^{\mi q z} \left( 3 \tanh^2 z - (q^2 + 1) - 3 \mi q \tanh z \right) \\
        f_{R q}^2 &= F_{R q}^2 \me^{\mi q z} \left( \tanh z - \mi q \right),
    \end{aligned} \\
\vdots \\
(\sigma_q^d)^2 &= \half (q^2 + d^2) m^2 && \Biggl\{
    \begin{aligned}
        f_{L q}^d &= \frac{1}{\sigma_n^d} \left( \frac{dm}{\sqrt2} \tanh z - \frac{\md}{\md w} \right) f_{L q}^{d-1} \\
        f_{R q}^d &= f_{L q}^{d-1}(w).
    \end{aligned}
\end{align*}
The normalisation coefficients $F$ can be computed using
Appendix~\ref{app:smpt}.

Using these basis functions, we expand the original
action~\eqref{eq:psi-act-5d} and integrate over the extra
dimension.  The effective four dimensional action, including the
kink dynamics, is
\begin{equation*}
\mcs_{\Phi+\Psi} = \int \md^4 x \left[ -\varepsilon_{\phi_c} + \mcl_\phi + \mcl_\psi \right],
\end{equation*}
where the Lagrangian for the expanded $\Psi$ is
\begin{equation}
\label{eq:psi-lag-4d}
\begin{aligned}
\mcl_\psi &=
    \overline{\psi_{L 0}} \mi \slashed\partial \psi_{L 0}
    + \sum_{i=1}^{\lceil d-1 \rceil} \overline{\psi_i} (\mi \slashed\partial - \sigma_i) \psi_i
    + \int_{-\infty}^{\infty} \md q \left[ \overline{\psi_q} (\mi \slashed\partial - \sigma_q) \psi_q \right] \\
& \quad
    - h_{i j k} \phi_i \overline{\psi_{L j}} \psi_{R k}
    - h_{i k j}^* \phi_i^* \overline{\psi_{R j}} \psi_{L k}.
\end{aligned}
\end{equation}
We have condensed the notation using
$\psi_i=\psi_{L i}+\psi_{R i}$ and similarly for $\psi_q$.  For
brevity in the Yukawa terms, the implicit sum over $i$ denotes a
sum over the bound modes $i=0,1$ and an integral over the
continuum; similarly the sum over $j$ and $k$ denotes the sum over
bound and continuum fermion modes.  The effective dimensionless
Yukawa coupling is
\begin{equation*}
h_{i j k} = \sqrt{\frac{a d^2}{2m}} \int_{-\infty}^\infty \md w \left[ \eta_i f_{L j}^* f_{R k} \right].
\end{equation*}
We can compute these $h$.  Those of importance are the
couplings between the bound kink modes and the bound fermion
modes, the first few being
\begin{align*}
h_{0 i 0} &= 0 \qquad \text{(for all $i$)}, \\
h_{0 0 1} &= \sqrt{\frac{3a}{8\sqrt2}}
             \left(\frac{d-\half}{d-1}\right)^{\thhalf}
             \frac{\Gamma^2(d-\half)}{\Gamma^2(d-1)}, \\
h_{0 1 1} &= h_{1 0 1} = 0, \\
h_{1 1 1} &= \sqrt{\frac{3a}{8\sqrt2}}
             \frac{(d-\half)^{\half}}{d-1}
             \frac{\Gamma^2(d-\half)}{\Gamma^2(d-1)}.
\end{align*}

This analysis includes the well known chiral zero mode
localisation when all fermion modes except $\psi_{L0}$ are removed
from~\eqref{eq:psi-lag-4d}.  In this reduced model there are no
Yukawa couplings between $\psi_{L0}$ and any of the kink modes
$\phi_i$ due to the chirality of the fermion mode.


\subsection{The thin kink with a fermion field}

With the expanded effective four dimensional action for $\Psi$, we
can now consider the limits of the kink.  With no kink ($\mu<0$) the basis
$f_i$ is not valid and we instead obtain a model containing the
coupled five dimensional fields $\Phi$ and $\Psi$.  The thick kink
scenario ($\mu=0$) contains four dimensional interacting scalar fields
$\phi_i$ and $\psi_i$.  The thin kink limit ($\mu>0$) is what we are most
interested in, where the remnant of the kink sector is just the
energy density.  Then the Yukawa term in~\eqref{eq:psi-act-5d}
reduces to $(dm/\sqrt2) \tanh^2 z \, \overline\Psi \Psi$ and we have
only the parameter $d$ left to play with.  There are four
scenarios, which we classify using the previous parameterisation
of the limit and write $d=\tilde d\Lambda^\delta$.

First, if $\delta=-\mu$ then $d\rightarrow0$, there are no bound
fermion modes and the fermion is left as a five dimensional field
with the action
\begin{equation*}
\mcs_{\Phi+\Psi}^\text{5D} =
    \int \md^4 x \left[ -\varepsilon_{\phi_c} \right]
    + \int \md^5 x \left[ \overline\Psi \mi \Gamma^M \partial_M \Psi
                          - \frac{dm}{\sqrt2} \theta(w) \overline\Psi \Psi \right],
\end{equation*}
where $\theta(w)$ is the step function.  The fermion in this case
has an unusual mass term that changes sign across the kink.
Second, if $\delta<-\mu$, we have the same situation as in the
first case, except the mass term disappears.  Third, if
$-\mu<\delta<0$, $d$ is not going to zero fast enough to counter
$m$ and the unusual mass term becomes infinite.  The five
dimensional field $\Psi$ is thus frozen out and we are left with
just the kink energy density in our effective theory.

The fourth case has $\delta\ge0$, so $d>0$ and there is at least
one bound fermion mode.  Because the masses of these modes go like
$m$, all modes are frozen out except the zero mode $\psi_{L0}$.
None of the Yukawa couplings $h$ are relevant because they couple
$\psi_{L0}$ to higher fermion modes which are not dynamical.
The effective action is then simply
\begin{equation*}
\mcs_{\Phi+\Psi}^\text{4D} =
    \int \md^4 x \left[ -\varepsilon_{\phi_c} + \overline{\psi_{L0}} \mi \slashed\partial \psi_{L0} \right],
\end{equation*}
which contains the dynamics of a single, four dimensional, chiral,
massless fermion.  Thus we obtain the result of Rubakov and
Shaposhnikov in the thin kink limit with $d>0$.  This is of
obvious importance in model building where chiral zero modes are
the starting point for the particle content.

In all the limits of the models we have considered so far, the
dynamical fields are either exclusively four dimensional or
exclusively five dimensional.  It is possible to construct an
action which in the thin kink limit contains interacting four
and five dimensional fields.  Because the dynamics of the kink
are frozen out, two extra fields are required from the outset: one
which couples appropriately to the kink and provides a four
dimensional zero-mode, and another which remains five dimensional.
To demonstrate this idea, we take the original action for the kink
and the fermion and add the scalar field $\Xi$ with a coupling to
the fermion only.  The action is then
\begin{equation*}
\mcs_\text{all} =
    \mcs_\Phi
    + \mcs_\Psi
    + \int \md^5 x \left[ \half \partial^M \Xi \partial_M \Xi - s \, \Xi (\overline\Psi \Psi^c + \overline{\Psi^c} \Psi) \right],
\end{equation*}
where $s$ is a dimensionful coupling constant and the charge
conjugate field is defined as $\Psi^c = \Gamma^2\Gamma^4\Psi^*$.
We choose the kink-fermion coupling $d>0$ such that there is at
least one bound mode and follow the thin kink analysis performed
above.  All the massive fermion modes are frozen out and the
$\Xi$ coupling term becomes
\begin{equation}
\label{eq:delta-couple}
s \, \Xi (\overline\Psi \Psi^c + \overline{\Psi^c} \Psi) \xrightarrow{m\rightarrow\infty} s \, (f_{L0})^2 \Xi (\overline{\psi_{L0}} \psi_{L0}^c + \overline{\psi_{L0}^c} \psi_{L0}).
\end{equation}
The extra dimensional factor from the fermion zero mode becomes a
delta distribution in the thin kink limit
\begin{equation*}
(f_{L0})^2 =
    \frac{m \, \Gamma(d+\half)}{\sqrt{2\pi} \, \Gamma(d)} \cosh^{-2d} \left( \frac{m w}{\sqrt2} \right)
    \xrightarrow{m\rightarrow\infty} \delta(w).
\end{equation*}
The $w$ integral can then be performed over the Yukawa term given
by~\eqref{eq:delta-couple} which reduces the action to
\begin{equation*}
\begin{aligned}
\mcs_\text{all}^\text{4D/5D} &=
    \int \md^4 x \left[
        -\varepsilon_{\phi_c}
        + \overline{\psi_{L0}} \mi \slashed\partial \psi_{L0} 
        - s \, \Xi(w=0) (\overline{\psi_{L0}} \psi_{L0}^c + \overline{\psi_{L0}^c} \psi_{L0}) \right] \\
&\quad
    + \int \md^5 x \left[
        \half \partial^M \Xi \partial_M \Xi \right].
\end{aligned}
\end{equation*}

This is a dynamically generated model describing a four dimensional
zero mode $\psi_{L0}$ coupled at the extra dimensional point $w=0$
to a five dimensional field $\Xi$.  The situation can easily be
reversed to have $\Psi$ five dimensional and coupling at $w=0$ to
the four dimensional ground state mode of $\Xi$.  Extensions to
multiple four and five dimensional fields are also easily
obtained.


\section{Conclusion}
\label{sec:conc}

Our universe may be embedded in a large extra spatial dimension
and it seems sensible to look for a field theoretic description of
such a scenario.  We would like such a model to explain the
dynamics behind the confinement of higher dimensional fields to
our four dimensional subspace.  We have demonstrated in this paper
that the domain wall defect (the kink) is an attractive candidate
for such dynamical confinement as it can localise not only
massless chiral fermions but also scalars with arbitrary quartic
potentials.

We have analysed in detail the kink and its modes and discussed
the behaviour of these degrees of freedom in the limits of no
kink, a thick kink and a thin kink.  Due to the quartic potential
which sets up the kink profile, the Kaluza-Klein spectrum consists
of a massless bound mode, a massive bound mode and a massive
continuum.  In the thin kink limit all the massive modes freeze
out as their mass becomes infinite.  The zero mode also freezes
out in this limit due to its divergent quartic self coupling.
This leaves just the kink energy density in the effective four
dimensional action.

A full analysis of a second scalar field coupled to the kink was
performed.  We showed that this field is trapped by the kink in a
symmetric modified P\"oschl-Teller potential well and we gave the
mass spectrum for this trapping.  In the thin kink limit, this
extra scalar field could either freeze out completely, be a free
five dimensional field, or have just its ground state mode in an
effective four dimensional action.  We showed that in this latter
case, an arbitrary quartic potential could be generated for this
mode and it could thus be used as a standard model electroweak
Higgs field.

Our final analysis was that of a fermion coupled to the kink.  It
has been known for some time that such a model admits a massless
chiral mode.  We generalised this result by giving the full
spectrum of fermion modes in the presence of the kink.  These
four dimensional fields consist of a left-handed zero mode, a
certain number of pairs of left- and right-handed modes and a
continuum of pairs.  In the thin kink limit, we showed that the
fermion can either freeze out completely, be five dimensional with
or without a mass term, or reproduce the result of a localised
four dimensional, massless, left-handed mode.  We also
demonstrated that with the kink and fermion and an extra scalar
field, one can arrange things such that the left-handed zero mode
couples to a five dimensional field at a single point in the extra
dimension.

The results of the present work gives one the tools necessary to
write down the standard model, without gauge fields, on a brane in
five dimensions.  With the details of the spectrum of the
decomposed five dimensional fields, one can compute the
phenomenology of the interactions between the massless and
massive modes and also use experimental data to put an upper bound
on the width of the brane.  Further work would focus on the
addition of gauge fields and their localisation, as well as the
inclusion of gravity.


\begin{acknowledgments}
We thank Mark Trodden for very fruitful discussions.
DPG was supported by the Puzey bequest to the University of
Melbourne and RRV by the Australian Research Council.
\end{acknowledgments}


\appendix

\numberwithin{equation}{section}

\section{Symmetric modified P\"oschl-Teller potential mode solutions}
\label{app:smpt}

In this appendix we give analytic solutions to the symmetric
modified P\"oschl-Teller potential.  The non-symmetric potential
was first studied by Rosen and Morse in the context of molecular
dynamics~\cite{rosen1932}.  They presented solutions in terms
of hypergeometric functions.  Later work by Nieto~\cite{nieto1978}
computed explicit forms of the bound mode solutions, including
normalisation coefficients, in terms of regular functions.
Rajaraman~\cite{rajaraman1982} gives unnormalised solutions for
the bound and continuum modes for a specific case of the
potential, and the hypergeometric forms of the solutions are
again explored in~\cite{hung2004}.  Bound state solutions are
also expressed in terms of Gegenbauer polynomials
in~\cite{decastro2006}.  As far as we are aware, exact closed form
solutions for the continuum modes and their normalisation factors
have not been presented in the literature.  We present these forms
in this Appendix, along with simple expressions for the bound
states, simple recurrence relations for higher modes,
normalisation coefficients, and the closure relation.

The time-independent Schr\"odinger equation with the symmetric
version of the potential takes the form
\begin{equation}
\label{eq:smpt}
\left( -\frac{\md^2}{\md x^2} + l(l+1) \tanh^2 x - l \right) \psi_n = E_n \psi_n.
\end{equation}
If $l=0$ then the solutions are just plane waves.  For $l>0$ there
are a set of bound modes followed by continuum modes.  The bound
solutions are
\begin{align*}
E_0^l &= 0 &
\psi_0^l(x) &= A_0^l \cosh^{-l} x, \\
E_1^l &= 2l-1 &
\psi_1^l(x) &= A_1^l \sinh x \cosh^{-l} x, \\
E_2^l &= 4l-4 &
\psi_2^l(x) &= A_2^l \left( \frac{2l-2}{2l-1} \cosh^{-l+2} x - \cosh^{-l} x \right), \\
\vdots \\
E_n^l &= 2nl-n^2 &
\psi_n^l(x) &= \frac{1}{\sqrt{E_n^l}} \left( l \tanh x - \frac{\md}{\md x} \right) \psi_{n-1}^{l-1}(x).
\end{align*}
The square integrable ortho-normalisation condition is
\begin{equation}
\label{eq:norm-sqint}
\int_{-\infty}^{\infty} \psi_n^l(x) \psi_{n'}^l(x) \; \md x = \delta_{n n'},
\end{equation}
and the normalisation coefficients are
\begin{align*}
A_0^l &= \sqrt\frac{\Gamma(l+\half)}{\sqrt{\pi} \, \Gamma(l)}, &
A_1^l &= \sqrt{2l-2} A_0^l, &
A_2^l &= \sqrt{(2l-1)(l-2)} A_0^l.
\end{align*}
These bound mode solutions are valid for for all positive real
values of $l$.  There are $\lceil l \rceil$ bound
modes\footnotemark[2] and so the mode index takes the values
$n = 0,1,\ldots,\lceil l-1 \rceil$.  For the continuum we have
found forms for the solutions in terms of regular functions for
the case where $l$ is a positive integer.  Instead of the discrete
bound mode index $n$, the continuum is indexed with a continuous
label $p\in\mbr$.  They take the form
\begin{align*}
E_p^1 &= p^2 + 1 &
\psi_p^1(x) &= A_p^1 \me^{\mi p x} \left( \tanh x - \mi p \right), \\
E_p^2 &= p^2 + 4 &
\psi_p^2(x) &= A_p^2 \me^{\mi p x} \left( 3 \tanh^2 x - (p^2 + 1) - 3 \mi p \tanh x \right), \\
\vdots \\
E_p^l &= p^2 + l^2 &
\psi_p^l(x) &= \frac{1}{\sqrt{E_p^l}} \left( l \tanh x - \frac{\md}{\md x} \right) \psi_p^{l-1}(x).
\end{align*}
The delta distribution ortho-normalisation condition is
\begin{equation}
\label{eq:norm-delta}
\int_{-\infty}^{\infty} \psi_p^l(x) \psi_{p'}^l(x)^* \; \md x = \delta(p-p'),
\end{equation}
and the normalisation coefficients are
\begin{align*}
A_p^1 &= \frac{1}{\sqrt{2\pi}} \frac{1}{\sqrt{p^2+1}}, &
A_p^2 &= \frac{1}{\sqrt{p^2+4}} A_p^1.
\end{align*}
These continuum modes are valid only for $l=1,2,3,...$; for other
values of $l$ one must resort to the hypergeometric form.  Note
that the bound modes are orthogonal to the continuum modes.
Finally, we state the closure relation:
\begin{equation*}
\sum_{n=0}^{\lceil l-1 \rceil} \psi_n^l(x) \psi_n^l(x')
  + \int_{-\infty}^{\infty} \psi_p^l(x) \psi_p^l(x')^* \; \md p = \delta(x-x').
\end{equation*}


\section{Freezing out of the kink translation dynamics}
\label{app:freeze}

The thin kink limit, as parameterised by the width going to zero
while keeping the energy density finite, still retains the
continuous translation symmetry of the original thick kink.  It is
unexpected then, that in this limit, the symmetry has no
observable consequence at the effective four dimensional level.
The physical reason for this is that a thin kink is infinitely
stiff and cannot be perturbed; the propagating scalar fields have
been frozen out.  With no dynamical degrees of freedom left, the
translation symmetry is manifest by the zero energy cost of a
specific constant field configuration which has no physical
consequence.  This constant configuration is equivalent to
shifting the mode expansion basis to align it with the translated
kink profile.  We discuss this in detail in the following
sections.

\subsection{No dynamical degrees of freedom}

We proceed to show that there are no effective four dimensional
degrees of freedom associated with a kink background in the thin
kink limit.  We do this by looking at the action for a general
field expanded about the kink, and show that the potential of such
a field has coefficients which tend to infinity in the thin kink
limit.

Because the kink modes $\eta_{0,1,q}$ derived in
Section~\ref{sec:kink} form a basis of extra dimensional profiles,
the corresponding fields $\phi_{0,1,q}$ form a basis of four
dimensional scalar degrees of freedom.  So any dynamical
behaviour of the kink that manifests itself in four dimensions as
a scalar field, must be some linear combination of the fields
$\phi_{0,1,q}$.  Let this arbitrary kink degree of freedom be
$\psi_0(x^\mu)$ so we can write
\begin{equation}
\label{eq:psi-comb}
\psi_0 = \alpha_0 \phi_0 + \beta_0 \phi_1 + \int \md q [\gamma_0^*(q) \phi_q].
\end{equation}

To determine the effective action for $\psi_0$ we will need to
invert Eq.~\eqref{eq:psi-comb} to express $\phi_{0,1,q}$ in
terms of $\psi_0$.  To do this properly, we need a complete
basis which includes $\psi_0$; call it $\psi_{0,n,q}$, where
$n$ labels the extra bound modes and $q$ labels the extra
continuum modes needed to complete the basis.  Then the inverse
of~\eqref{eq:psi-comb} is
\begin{align*}
\phi_0 &= \alpha_0 \psi_0
    + \sum_n \alpha_n \psi_n
    + \int \md r [ \alpha_r \psi_r ], \\
\phi_1 &= \beta_0 \psi_0
    + \sum_n \beta_n \psi_n
    + \int \md r [ \beta_r \psi_r ], \\
\phi_q &= \gamma_0(q) \psi_0
    + \sum_n \gamma_n(q) \psi_n
    + \int \md r [ \gamma_r(q) \psi_r ].
\end{align*}
Substituting this expansion in the action~\eqref{eq:phi-act-4d}
gives the full action for the new fields $\psi_{0,n,q}$.  We are
only interested in the terms that describe $\psi_0$ as a
self-interacting field that does not couple to any of the other
fields.  Keeping just these terms, the effective action is
\begin{align*}
S_{\psi_0} = \int \md^4 x \left[
    \half A^\text{(kin)} \partial^\mu \psi_0 \partial_\mu \psi_0
    - \frac{m^2}{2} A^{(2)} \psi_0^2
    - m \sqrt{a} \, A^{(3)} \psi_0^3 - a A^{(4)} \psi_0^4 \right],
\end{align*}
where
\begin{align*}
A^\text{(kin)} &= \alpha_0^2 + \beta_0^2 + \int \md q \left[ \gamma_0^*(q) \gamma_0(q) \right], \\
A^{(2)} &= \thhalf \beta_0^2 + \int \md q \left[ \half (q^2 + 4) \gamma_0^*(q) \gamma_0(q) \right], \\
A^{(3)} &= \text{(terms with at least one factor of $\beta_0$ or $\gamma_0(q)$)}, \\
A^{(4)} &= \frac{9 \sqrt2}{140} \alpha_0^4 + \text{(terms with at least one factor of $\beta_0$ or $\gamma_0(q)$)}.
\end{align*}

In the thin kink limit\footnote{Recall the limit is
$m\rightarrow\infty$ and $a\rightarrow\infty$  but $m^4/a$
finite.} the mass of $\psi_0$ will tend to infinity unless
$A^{(2)}$ vanishes.
For $A^{(2)}$ to be identically zero, we require $\beta_0 = 0$ and
$\gamma_0(q) = 0$.  Then because $A^\text{(kin)}=1$, we require
$\alpha_0 = 1$ (hence $\psi_0 = \phi_0$) and so the quartic
coefficient is $9 \sqrt2 \, a / 140$ which freezes out $\psi_0$ in
the thin kink limit.  Alternatively, we can choose $\beta_0$
and $\gamma_0(q)$ such that $A^{(2)}$ is not identically zero, but
in the thin kink limit tends to zero in order to keep
$m^2 A^{(2)}$ finite.  Then in this limit
$\beta_0 \rightarrow 0$ and $\gamma_0(q) \rightarrow 0$ and
so $\alpha_0 \rightarrow 1$.  The quartic coefficient $A^{(4)}$ is
then dominated by the $\alpha_0^4$ term which, as before, tends to
infinity in the thin kink limit and freezes out $\psi_0$.  Thus
there is no linear combination of the fields $\phi_{0,1,q}$, and
hence no scalar degree of freedom, whose dynamics survive in the
thin kink limit.

\subsection{Manifestation of translation symmetry}

The energy density $\varepsilon_{\phi_c}$ of the kink
configuration $\phi_c(w)$ is independent of the location of the
kink along the $w$-dimension.  Thus, for a linear combination of
the basis $\eta_{0,1,q}$ which just translates the kink profile,
the corresponding linear combination of $\phi_{0,1,q}$ should have
zero energy density.  This statement needs to be made more
precise, as the previous section shows that there is no linear
combination of $\phi_{0,1,q}$ that has a vanishing potential.
Instead, there is a linear combination with a potential that
vanishes only at a specific value of the field, corresponding to a
shift of the kink.

Consider the expansion
\begin{equation}
\label{eq:phi-expand-shift}
\Phi(x^M) = \phi_c(w) + \psi_0(x^\mu) s_0(w),
\end{equation}
where $s_0$ is the profile of a finite shift of the kink.  For a
given fixed shift of the kink $\Delta w$, we want to have the
relation
\begin{equation}
\label{eq:shift-defn}
\psi_0(x^\mu) s_0(w) = \phi_c(w+\Delta w) - \phi_c(w).
\end{equation}
For this to be true, $\psi_0$ must be an arbitrary constant,
which we fix by normalising the profile
$\int \md w \, s_0^2(w) = 1$.  This gives
\begin{equation*}
s_0(w) = \sqrt{m} \, S \sigma_0(w),
\end{equation*}
with
\begin{align*}
& S = \frac{1}{2^{5/4}} \sqrt{\frac{\tanh \Delta z}{\Delta z - \tanh \Delta z}}
& \text{and}
&& \sigma_0(w) = \frac{\tanh \Delta z \, \cosh^{-2} z}{1 + \tanh \Delta z \, \tanh z}.
\end{align*}
We have $z = mw/\sqrt2$ and $\Delta z = m \Delta w / \sqrt2$. For
the combination $\psi_0 s_0$ to actually shift $\phi_c$ by an
amount $\Delta w$, as given by~\eqref{eq:shift-defn}, it is
required that
\begin{equation}
\label{eq:shift-psi}
\psi_0(x^\mu) = \frac{m}{\sqrt a \, S}.
\end{equation}

We now want to treat the field $\psi_0$ as a dynamical four
dimensional scalar degree of freedom.  It will correspond to
finite translations of the kink profile and we are interested in
its behaviour in the thin kink limit.  One can write $\psi_0$ as
a linear combination of $\phi_{0,1,q}$, which can be determined
from the linear combination needed to write $s_0$ in terms of
$\eta_{0,1,q}$.  It is then possible to determine the effective
action for $\psi_0$ by substituting this combination in the
Lagrangian~\eqref{eq:phi-lag-4d}.  A simpler way to obtain the
same result is to substitute~\eqref{eq:phi-expand-shift} into the
original five dimensional action~\eqref{eq:phi-act-5d} and
integrate out the extra dimension.  The result is
\begin{equation*}
\mcs_{\psi_0} = \int \md^4 x \left[
    -\varepsilon_{\phi_c} 
    + \half \partial^\mu \psi_0 \partial_\mu \psi_0
    - V_{\psi_0} (\psi_0) \right],
\end{equation*}
with the effective potential
\begin{equation*}
V_{\psi_0} (\psi_0) = \left( \frac{5}{2 \tanh^2 \Delta z} - \frac{5 \tanh \Delta z}{6 (\Delta z - \tanh \Delta z)} - \frac{3}{2} \right) \psi_0^2 \left( \sqrt{a} S \psi_0 - m \right)^2.
\end{equation*}

Note that the potential for $\psi_0$ has two minima: one at
$\psi_0=0$ and the other at precisely the value corresponding to
the finite translation, given by~\eqref{eq:shift-psi}.  The
energy density associated with $\psi_0$ is
\begin{equation}
\label{eq:shift-psi-energy}
\varepsilon_{\psi_0} = \half (\partial_t \psi_0)^2 + \half (\nabla \psi_0)^2 + V_{\psi_0}(\psi_0).
\end{equation}
Any dynamical behaviour of $\psi_0$ has non-zero, positive energy
density.  This energy density will vanish only when $\psi_0=0$ or
$\psi_0$ is the constant~\eqref{eq:shift-psi} at all locations in
the four dimensional space-time.  This is the manifestation of the
translation symmetry of the kink.  The field $\psi_0$, or
equivalently a specific linear combination of $\phi_{0,1,q}$,
can assume this constant value at each point $x^\mu$ with zero
energy cost.  It is possible to show that the fields assuming this
value is equivalent to shifting the original basis $\eta_{0,1,q}$
by the amount $\Delta w$.

In the thin kink limit, the potential $V_{\psi_0}$ becomes
infinitely steep and the dynamics of $\psi_0$ are frozen out.  The
only remnant of the translation symmetry is the fact that $\psi_0$
is allowed to be a certain constant value everywhere with no
energy cost.  But this has no effect on the four dimensional
effective action and so the translation symmetry is hidden at this
level.



\end{document}